\documentclass[aps,pre,reprint,superscriptaddress,nofootinbib,floatfix,twocolumn]{revtex4-1}

\usepackage{amsmath,amssymb,amsfonts,bm}
\usepackage{graphicx}
\usepackage{dcolumn}
\usepackage{xcolor}
\usepackage{placeins}
\definecolor{colorFC}{HTML}{1F77B4}  
\definecolor{colorGC}{HTML}{D62728}  
\definecolor{colorHC}{HTML}{2CA02C}  
\definecolor{colorCRW}{HTML}{555555} 
\definecolor{colorSigmaT}{HTML}{DC143C}  

\usepackage{hyperref}
\usepackage{subcaption}
\hypersetup{
    colorlinks=true,
    linkcolor=blue,
    citecolor=blue,
    urlcolor=blue
}

\begin{document}

\title{Interference-induced dynamical crossovers in quantum walks on disordered square lattices}

\author{\textbf{Ramnarayan Bera}} 
\email{ramnarayan97@jcbose.ac.in}  
\author{\textbf{Arijit Roy}}
\email{arijitroy@jcbose.ac.in}
\author{\textbf{ Soumyadip Ghosh}}
\email{soumyadip.ghosh@jcbose.ac.in}
\author{\textbf{ Soumen Roy}}
\email{soumen@jcbose.ac.in}
\affiliation{Department of Physical Sciences, Bose Institute, Kolkata 700091, India}


\begin{abstract}
{We study discrete-time quantum walks with Fourier, Grover, and Hadamard coin operators on disordered square lattices, alongside classical random walks.  Edges in the lattice are absent with a probability, $p$. In this case, both classical and quantum walks exhibit an essentially identical degradation of transport. We then study a general version of this model, in which the existing edges are retained. However, the absent edges are no longer absent, but become partially transmitting, with a transmission coefficient $f$. We find that the response is now quite sensitive to the walk protocol and coin symmetry. The classical walk exhibits asymptotic stability of the anomalous diffusion exponent, $\beta(t)$. In comparison, quantum walks exhibit a monotonically decaying $\beta(t)$ beyond the initial transient regime, leading to the absence of quantum diffusion in two dimensions. We map the transport regimes in the $(f,p)$ plane and identify the contour for $\beta=1$, which separates the subdiffusive and superdiffusive regimes. We also examine the effective diffusion coefficient, the inverse participation ratio, and the first two moments of edge transmission. We restrict ourselves to analyzing bulk propagation prior to finite-size boundary reflections. The results indicate that quantum transport depends significantly on coin-specific interference and cannot be fully described by transmission statistics alone.}
\end{abstract}

\maketitle

\section{Introduction}
\label{sec:intro}
 Quantum walks are the quantum-mechanical analogues of classical random walks and represent a foundational basis in quantum information science and quantum simulation~\cite{aharonov1993quantum, kempe2003quantum}. On translationally invariant lattices, DTQWs spread ballistically because of unitary evolution and quantum interference. Their mean-squared displacement scales as $\sigma^2(t)\sim t^\beta$, with $\beta=2$. By contrast, a CRW shows diffusive scaling with $\beta=1$. Quantum walks also provide a framework for studying quantum search, spatial exploration on complex networks, and state-transfer protocols~\cite{childs2004spatial, venegas2012quantum}.
 
 Disorder can disrupt coherent quantum transport in crystalline and networked systems. Impurities, broken links, and phase fluctuations alter propagation. The scaling theory of localization predicts Anderson localization for non-interacting particles in two dimensions with static disorder~\cite{abrahams1979scaling,anderson1958absence}. This prediction concerns the infinite-system limit and the assumptions of the scaling theory. In a bond-percolation model, each edge is removed with probability $p$. Edge removal can break network connectivity and create bottlenecks. These geometric effects may dominate transport and obscure differences between quantum interference and classical diffusion.

To separate geometric disconnection from coherent scattering, we also study a beam-splitter disorder model. In this model, defects have transmission coefficient $0\leq f\leq1$ and occur with probability $p$. Unlike bond removal, partial transmission preserves the links while allowing reflection and scattering. The transport therefore depends on both the disorder parameters and the coin operator. We compare the time dependence of $\beta(t)$ for classical and quantum walks to test whether the quantum walk shows a sustained diffusive regime within the simulated time window.

 We compare DTQWs and CRWs on disordered two-dimensional square lattices. We consider Fourier, Grover, and Hadamard coin operators. First, we examine transport under binary bond percolation. In this model, edges are either present or absent. We then study the beam-splitter model, which allows partial transmission and reflection. We compare the transport response across the walk protocols and coin operators.Our main findings are as follows:\\
{\tiny $\bullet$} {Absence of stable quantum diffusion:} We find that the classical exponent $\beta(t)$ approaches a plateau, whereas the quantum exponent decreases over the analysed time window. This behaviour is consistent with localization, but the finite-time data alone do not prove asymptotic localization.\\
{\tiny $\bullet$} {Non-equilibrium phase diagrams:} We map the measured transport regimes in the $(f,p)$ plane. The contour $\beta=1$ separates superdiffusive spreading ($\beta>1$) from subdiffusive spreading ($\beta<1$). These labels describe the observed scaling and do not, by themselves, identify a unique microscopic interference mechanism.\\
{\tiny $\bullet$}  {Effect of heterogeneity or percolation:} The extracted threshold $f_c(p)$ reaches a maximum near $p_c\approx0.50$ in our data. We examine whether this feature coincides with the bond-percolation threshold. A claim that it results from maximal destructive interference requires additional evidence.\\
{\tiny $\bullet$}  {Statistical transport diagnostics:} We analyse the early-time effective diffusion coefficient $D_{\mathrm{eff}}$, the inverse participation ratio (IPR), and the first two moments of link transmission. We compare these observables for classical and quantum walks. The results test whether transmission statistics describe the classical response and whether additional coin-dependent effects are needed to describe quantum transport.

The remainder of this paper is structured as follows. Section~\ref{sec:model} defines the quantum-walk protocol, the coin operators, and the two disorder models. Section~\ref{sec:diagnostics} introduces the transport observables, including $\beta(t)$, $D_{\mathrm{eff}}$, the IPR, and the transmission moments. Section~\ref{sec:results} presents the transport results, phase diagrams, localization indicators, and finite-size analysis. Section~\ref{sec:conclusion} summarizes the main findings and discusses possible experimental tests like photonic and quantum walk platforms.

\section{Theoretical Models and Disorder Formalisms}
\label{sec:model}

\subsection{The Two-Dimensional Discrete-Time Quantum Walk}

We consider a quantum walker on a finite $L\times L$ square
lattice. Each site is labelled by $\mathbf{x}=(x,y)$, where
$x,y\in\{0,\ldots,L-1\}$. The Hilbert space is
$\mathcal{H}=\mathcal{H}_c\otimes\mathcal{H}_p$, where
$\mathcal{H}_c$ is the four-dimensional coin space and
$\mathcal{H}_p$ is the position space. The directional coin
basis is
\begin{equation}
|0\rangle=|+x\rangle,\quad
|1\rangle=|-x\rangle,\quad
|2\rangle=|+y\rangle,\quad
|3\rangle=|-y\rangle.
\end{equation}
The state after $t$ steps is
\begin{equation}
|\psi(t)\rangle=
\sum_{d=0}^{3}\sum_{x,y=0}^{L-1}
\psi(d,x,y,t)|d\rangle\otimes|x,y\rangle.
\end{equation}
Each step consists of a coin operation followed by spatial
propagation:
\begin{equation}
|\psi(t+1)\rangle=
\hat{S}(\hat{C}\otimes\hat{I})|\psi(t)\rangle,
\label{eq:quantum_evolution}
\end{equation}
where $\hat{C}$ acts on the coin space and $\hat{S}$ includes
propagation and scattering at the lattice bonds. Coin-based
quantum walks and their generalizations are reviewed in
Refs.~\cite{kempe2003quantum,venegas2012quantum}.

\subsection{Coin Operators}

We compare three unitary coin operators. The four-dimensional
Fourier coin is
\begin{equation}
(C_F)_{jk}=\frac{1}{2}
\exp\left(\frac{2\pi i jk}{4}\right),
\qquad j,k\in\{0,1,2,3\}.
\label{eq:fourier_coin}
\end{equation}
The Grover coin is
\begin{equation}
C_G=\frac{1}{2}
\begin{pmatrix}
-1&1&1&1\\
1&-1&1&1\\
1&1&-1&1\\
1&1&1&-1
\end{pmatrix}.
\label{eq:grover_coin}
\end{equation}
The Hadamard tensor coin is
\begin{equation}
C_H=H_2\otimes H_2
=\frac{1}{2}
\begin{pmatrix}
1&1&1&1\\
1&-1&1&-1\\
1&1&-1&-1\\
1&-1&-1&1
\end{pmatrix},
\label{eq:hadamard_coin}
\end{equation}
where
\begin{equation}
H_2=\frac{1}{\sqrt{2}}
\begin{pmatrix}
1&1\\
1&-1
\end{pmatrix}.
\end{equation}

\subsection{Disorder in Bond Transmission}

We assign each nearest-neighbour bond an independent binary
disorder variable $\eta_e\in\{0,1\}$, with
\begin{equation}
\Pr(\eta_e=1)=p,\qquad
\Pr(\eta_e=0)=1-p.
\label{eq:bond_disorder}
\end{equation}
Here, $\eta_e=1$ denotes a defective bond, while $\eta_e=0$
denotes a pristine bond. Pristine bonds transmit perfectly.
Every defective bond has the same transmission probability
$f\in[0,1]$. Thus, $p$ controls the defect concentration, while
$f$ controls the transmission through each defect. This
formulation describes static bond disorder, as used in studies
of quantum walks on disordered lattices
\cite{chandrashekar2014disorder}.

For a pristine bond, the two counter-propagating amplitudes
pass through unchanged:
\begin{equation}
\begin{pmatrix}
\psi_{\mathrm{out}}^{(R)}\\
\psi_{\mathrm{out}}^{(L)}
\end{pmatrix}
=
\begin{pmatrix}
1&0\\
0&1
\end{pmatrix}
\begin{pmatrix}
\psi_{\mathrm{in}}^{(R)}\\
\psi_{\mathrm{in}}^{(L)}
\end{pmatrix}.
\end{equation}
For a defective bond, the local scattering rule is
\begin{equation}
U_f=
\begin{pmatrix}
\sqrt{f}&i\sqrt{1-f}\\
i\sqrt{1-f}&\sqrt{f}
\end{pmatrix},
\qquad U_f^\dagger U_f=I.
\label{eq:bond_scattering}
\end{equation}
The diagonal amplitudes describe transmission, while the
off-diagonal amplitudes describe reversal of propagation.
Their squared magnitudes are $f$ and $1-f$, respectively.
The relative phase $i$ ensures unitary scattering.

For a horizontal bond connecting $(x,y)$ and $(x+1,y)$,
let $a=\psi(0,x,y,t)$ and $b=\psi(1,x+1,y,t)$ denote the
incident amplitudes. For a defective bond, the outgoing
amplitudes are
\begin{align}
\psi_{\mathrm{new}}(0,x+1,y)
&=\sqrt{f}\,a+i\sqrt{1-f}\,b,
\\
\psi_{\mathrm{new}}(1,x,y)
&=i\sqrt{1-f}\,a+\sqrt{f}\,b.
\label{eq:horizontal_scattering}
\end{align}
Vertical bonds follow the same rule for the channels
$|+y\rangle$ and $|-y\rangle$. The bond variables remain fixed
during a walk, so the disorder is quenched.

\subsection{Classical Markov Model}

We construct a classical random walk on the same lattice and
directional channels. Its state is the probability vector
$\mathbf{P}(\mathbf{x},t)
=[P_0,P_1,P_2,P_3]^T$, normalized as
\begin{equation}
\sum_{x,y}\sum_{d=0}^{3}P_d(x,y,t)=1.
\end{equation}
The coin operation is represented by the doubly stochastic
matrix
\begin{equation}
T_{jk}=|C_{jk}|^2,\qquad
\sum_jT_{jk}=\sum_kT_{jk}=1.
\label{eq:classical_coin}
\end{equation}
For each of the three coins defined above, $T_{jk}=1/4$.
The classical coin therefore assigns equal transition
probabilities to all four directions.

Across a pristine bond, probability propagates without
mixing the counter-propagating channels. Across a defective
bond, the probability fluxes mix incoherently. For a
horizontal bond, with $a=P_0(x,y,t)$ and
$b=P_1(x+1,y,t)$, the update is
\begin{align}
P_{\mathrm{new}}(0,x+1,y)
&=fa+(1-f)b,
\\
P_{\mathrm{new}}(1,x,y)
&=(1-f)a+fb.
\label{eq:classical_bond}
\end{align}
The same rule applies to vertical bonds. These updates
conserve total probability and use the same bond parameters
as the quantum model. The classical dynamics contains no
coherent interference between amplitudes.

\section{Transport Observables and Dynamical Diagnostics}
\label{sec:diagnostics}

\subsection{Mean-Squared Displacement and Transport Exponent}
\label{subsec:diffusion_exponent}

Summing over the coin states gives the position probability
\begin{equation}
P(x,y,t)=
\begin{cases}
\displaystyle\sum_{c=0}^{3}
|\langle c,x,y|\psi(t)\rangle|^2,
& \text{quantum walk},\\[6pt]
\displaystyle\sum_{d=0}^{3}P_d(x,y,t),
& \text{classical walk}.
\end{cases}
\end{equation}
The mean-squared displacement (MSD), measured from the initial
position $(x_0,y_0)$, is
\begin{equation}
\sigma^2(t)=
\sum_{x,y}
\left[(x-x_0)^2+(y-y_0)^2\right]P(x,y,t).
\label{eq:msd}
\end{equation}
We characterize its growth using the local transport exponent
\cite{YaoWald2023}
\begin{equation}
\beta(t)=\frac{d\ln \sigma^2(t)}{d\ln t}.
\label{eq:beta_inst}
\end{equation}
The values $\beta=2$ and $\beta=1$ correspond to ballistic
spreading and normal diffusion, respectively. Values
$0<\beta<1$ indicate subdiffusive growth. And values like $1<\beta<2$ indicate superdiffusive growth.

In our calculations, the ensemble-averaged exponent
is calculated as
\begin{equation}
\beta = \frac{1}{N_r}\sum_{i=1}^{N_r}\beta_i,
\end{equation}
and the SEM is given by
\begin{equation}
\mathrm{SEM} = \frac{\sigma_\beta}{\sqrt{N_r}},
\end{equation}
where $\sigma_\beta$ is the sample standard deviation of the fitted
exponents and $N_r$ is the total number of independent disorder
realizations for a fixed lattice square lattice size with specified (f,p) value. The error bars thus quantify the statistical uncertainty
in the ensemble-averaged transport exponent arising from variations
across disorder realizations. This procedure is applied to both the
classical random walk and the quantum discrete-time walks.
 
\subsection{Effective Diffusion Coefficient}
\label{subsec:diff_coeff}

Over a time interval $[t_1,t_2]$, we define
\begin{equation}
D_{\mathrm{eff}}=
\frac{\sigma^2(t_2)-\sigma^2(t_1)}
{4(t_2-t_1)}.
\label{eq:deff}
\end{equation}
The factor of four follows from the two-dimensional diffusion
law $\sigma^2(t)\sim 4Dt$. We evaluate this quantity within the
bulk propagation window.

\subsection{Inverse Participation Ratio}
\label{subsec:ipr}

The spatial concentration of the walker is measured by the
inverse participation ratio \cite{Murphy2011}
\begin{equation}
\mathrm{IPR}(t)=\sum_{x,y}P(x,y,t)^2.
\label{eq:ipr}
\end{equation}
Larger values indicate a more concentrated distribution, while
smaller values indicate broader spreading. Its time dependence
provides a complementary diagnostic of localization
\cite{Schreiber2011,YaoWald2023}.

\subsection{Bulk Propagation Window}
\label{subsec:bulk_window}

We restrict the analysis to times before the walker can reach
a lattice boundary. If $d_{\mathrm{b}}$ is the distance from
the initial site to the nearest boundary and $v_{\max}$ is the
maximum propagation speed in lattice sites per time step, we use
\begin{equation}
t<t_{\mathrm{bound}}=\frac{d_{\mathrm{b}}}{v_{\max}}.
\label{eq:boundary_cutoff}
\end{equation}
For a walker starting at the centre of an $L\times L$ lattice,
$d_{\mathrm{b}}=(L-1)/2$. The cutoff must be adjusted if the
initial site or boundary conditions differ.

\subsection{Transmission Statistics}
\label{subsec:transmission_stats}

Let $T_e$ denote the transmission coefficient of link $e$. For a good edge/link $T=1$ and for a distorted or partially transmitting edge, $T=f$.
For $N_{\mathrm{edges}}$ links, the sample mean and variance are
\begin{equation}
\langle T\rangle=
\frac{1}{N_{\mathrm{edges}}}\sum_e T_e,
\qquad
\sigma_T^2=
\frac{1}{N_{\mathrm{edges}}}
\sum_e(T_e-\langle T\rangle)^2.
\label{eq:transmission_stats}
\end{equation}
If each link is pristine with probability $1-p$ and defective
with probability $p$, with respective transmissions $1$ and $f$,
the ensemble values are
\begin{equation}
\mathbb{E}[T]=1-p(1-f),
\qquad
\operatorname{Var}(T)=p(1-p)(1-f)^2.
\label{eq:transmission_theory}
\end{equation}

\section{Results and Discussion}
\label{sec:results}

We compare discrete-time quantum walks (DTQWs) and classical random walks
(CRWs) on disordered square lattices, considering the Fourier, Grover, and
Hadamard coins. We investigate how the defect probability $p$ and the
transmission probability $f$ affect transport, spatial concentration,
and finite-size dependence. We first examine the transport regimes across
parameter space, then compare the transport response with the statistical
properties of the bonds, and finally investigate the time dependence and
finite-size robustness of the observed dynamics.

\subsection{Transport Dynamics Across Disorder}
\label{subsec:weak_disorder_transport}

\begin{figure*}[!t]
    \centering
    \includegraphics[width=0.94\textwidth,height=0.42\textheight,keepaspectratio]{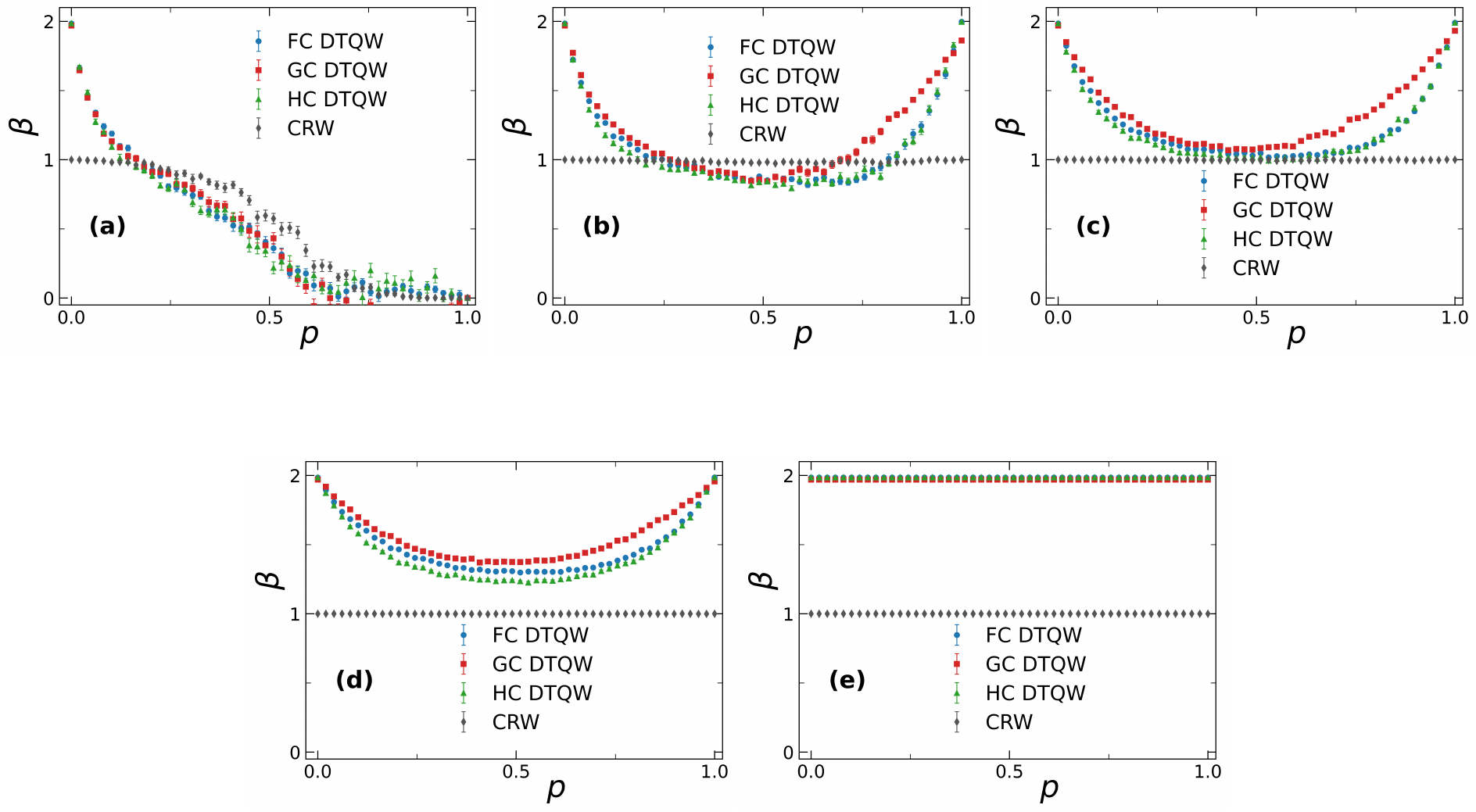}
    \caption{Transport exponent $\beta$ versus defect probability $p$ for edges with
different transmission coefficients $f$: (a) $f=0$, (b) $f=0.25$,
(c) $f=0.50$, (d) $f=0.75$, and (e) $f=1.00$. The panels compare
classical random walks (CRWs) with discrete-time quantum walks (DTQWs)
using the Fourier (FC), Grover (GC), and Hadamard (HC) coin operators.}
    \label{fig:weak_link_transport}
\end{figure*}

Figure~\ref{fig:weak_link_transport} compares the transport exponent as
the defect probability $p$ varies for different transmission probabilities
$f$. The classical and quantum walks exhibit different responses to the
disorder parameters, while the quantum results also depend on the coin
operator. These comparisons show that the transport response cannot be
characterized by the mean transmission alone. The dependence on both $p$
and $f$ motivates a systematic examination of the transport regimes across
the parameter plane.

\subsection{Absence of Sustained Quantum Diffusion}
\label{subsec:absence_quantum_diffusion}

\begin{figure*}[!t]
    \centering
    \includegraphics[width=0.90\textwidth,height=0.43\textheight,keepaspectratio]{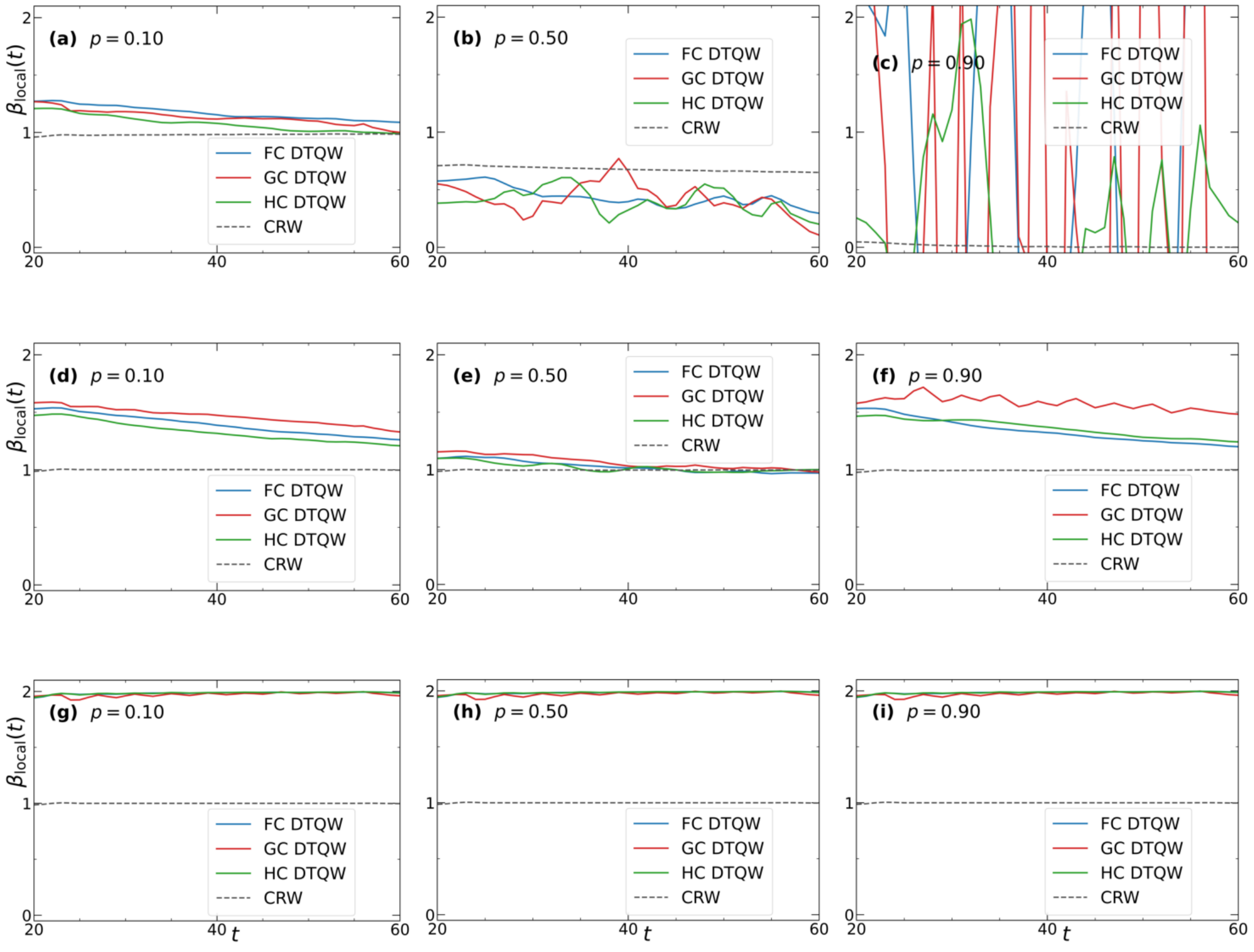}
    \caption{Temporal evolution of the local transport exponent
$\beta_{\mathrm{local}}(t)$
for different defect probabilities $p$ and transmission probabilities
$f$. Panels (a)--(c) correspond to $f=0$, (d)--(f) to $f=0.50$,
and (g)--(i) to $f=1.00$ with defective edge probability written inside graphs. The curves compare quantum walks
with Fourier (FC), Grover (GC), and Hadamard (HC) coins with the
classical random walk (CRW).}
    \label{fig:beta_matrix_9panel}
\end{figure*}

Figure~\ref{fig:beta_matrix_9panel} shows the time evolution of the
local transport exponent $\beta_{\mathrm{local}}(t)$ for different
values of $p$ and $f$. For all displayed defect probabilities, the
quantum transport exponent generally decreases with time when $f<1$,
although its evolution depends on the coin operator and disorder
parameters. In contrast, for $f=1$, the exponent remains close to the
ballistic value $\beta=2$, since all bonds transmit perfectly.
\\The classical exponent remains approximately stable over the simulated
time window. This difference indicates that partial transmission and
reflection affect quantum transport differently from classical
transport. The decay of $\beta_{\mathrm{local}}(t)$ suggests the
absence of sustained quantum diffusion in the simulated regime.
However, longer-time simulations are needed to establish the
asymptotic behaviour and confirm dynamical localization.

\subsection{Transport Regimes in the $(f,p)$ Plane}
\label{subsec:phase_diagrams}

\begin{figure*}[!t]
    \centering
    \includegraphics[width=0.94\textwidth,height=0.42\textheight,keepaspectratio]{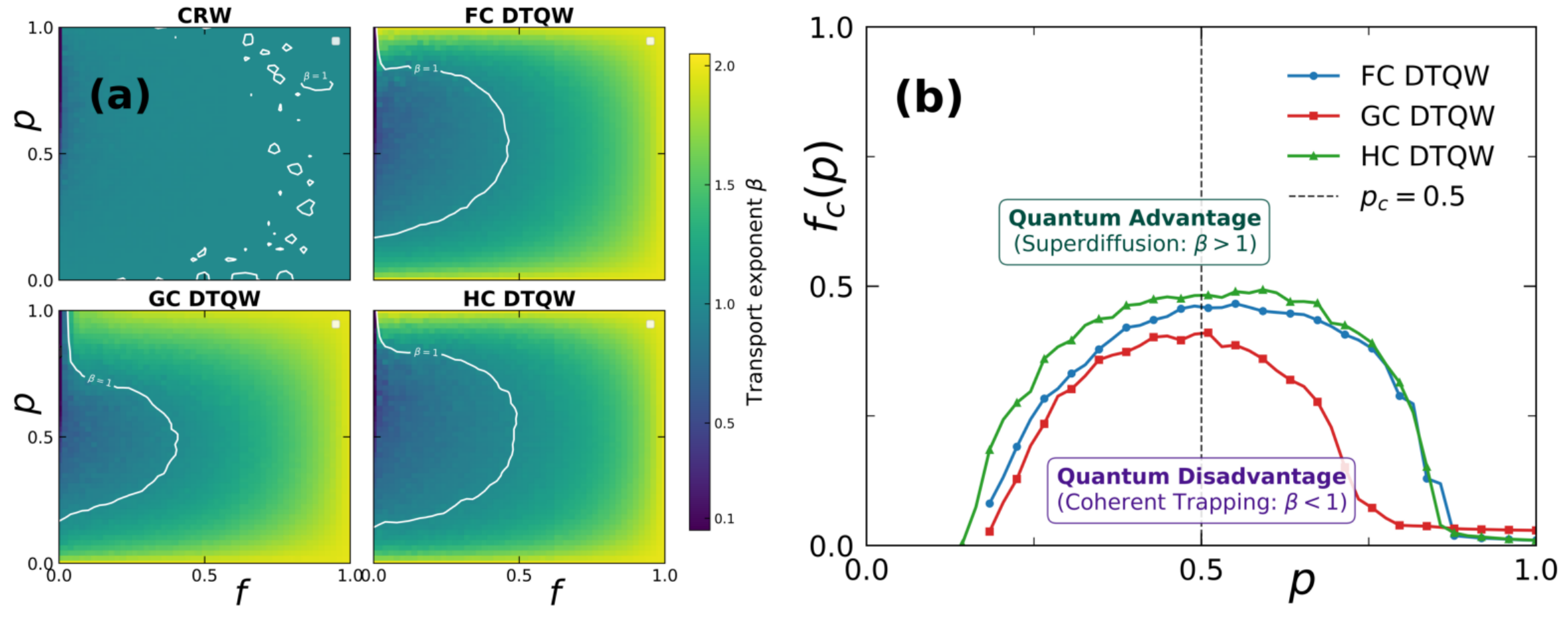}
    \caption{Transport-regime maps in the $(f,p)$ plane. (a) Regions
    classified by the measured exponent: $\beta<1$ indicates subdiffusive
    and $\beta>1$ superdiffusive spreading. The contour $\beta=1$ marks
    the crossover within the fitting window. (b) The corresponding
    $\beta=1$ contours for the Fourier, Grover, and Hadamard coins.}
    \label{fig:phase_diagrams}
\end{figure*}

Figure~\ref{fig:phase_diagrams} summarizes the measured transport regimes
in the $(f,p)$ plane. The regions with $\beta<1$ and $\beta>1$ correspond
to subdiffusive and superdiffusive spreading, respectively, within the
fitting window used to estimate the exponent. 
In the superdiffusive regime,
net constructive interference dominates, preserving a transient quantum speedup where spreading remains faster than classical Markovian diffusion. In sub-diffusive domain destructive interference and coherent back-scattering dominate, retarding spatial transport below the classical baseline.\\

The contour $\beta=1$
identifies the boundary between these two classifications. Its dependence
on the coin operator indicates that the same disorder parameters can
produce different transport responses for different quantum walks.
We therefore interpret this contour as a finite-time dynamical crossover,
rather than evidence of a thermodynamic phase transition.

\subsection{Transmission Statistics and Transport Response}
\label{subsec:transmission_moments}
\begin{figure*}[!t]
    \centering
    \includegraphics[width=0.94\textwidth,height=0.42\textheight,keepaspectratio]{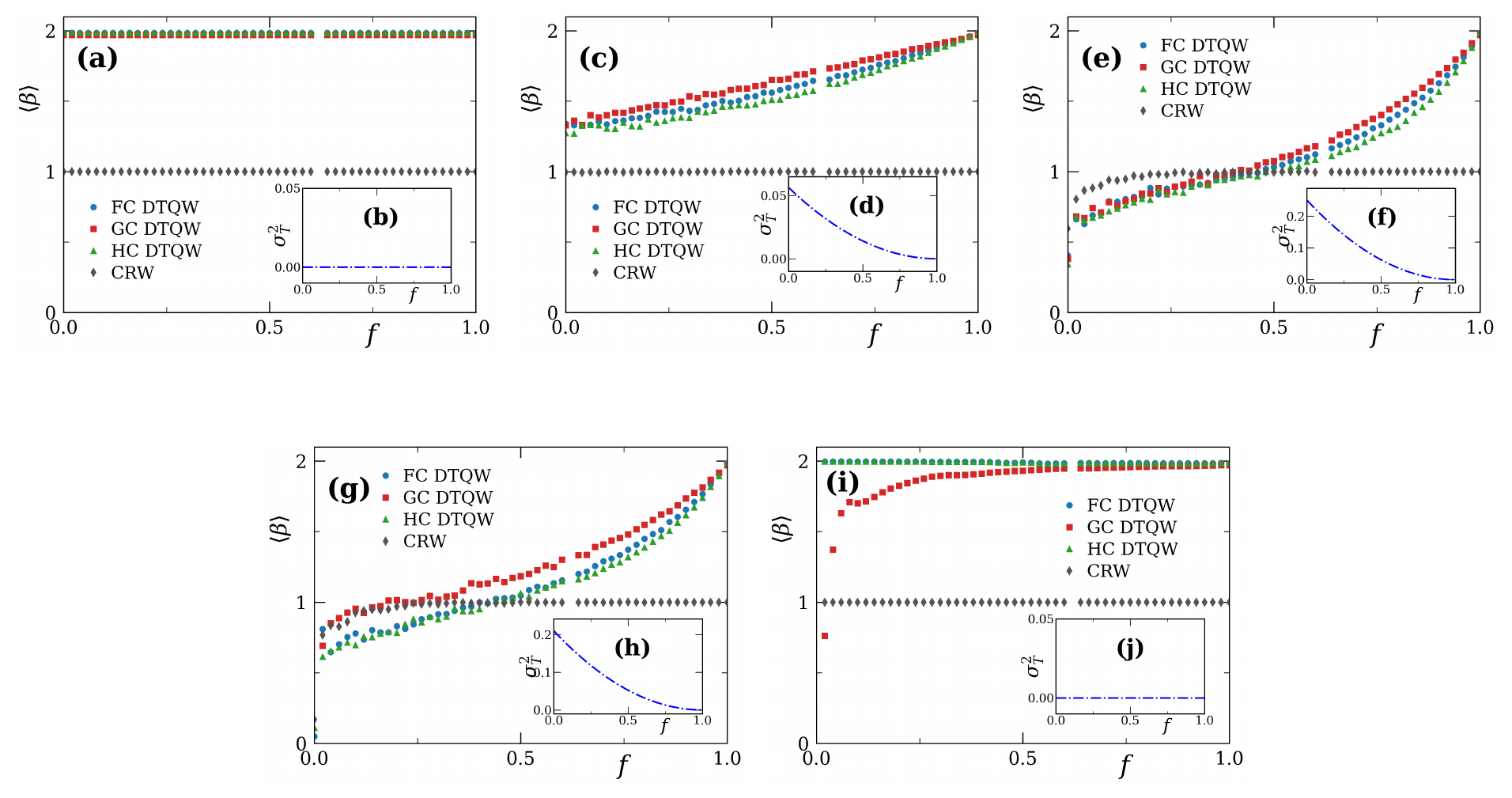}
    \caption{Mean transport exponent $\langle\beta\rangle$ as a function
of the transmission probability $f$ for defect probabilities
$p=0$, $0.06$, $0.50$, $0.70$, and $1.00$ in panels (a), (c), (e),
(g), and (i), respectively. The corresponding panels (b), (d), (f),
(h), and (j) show the variance of bond transmission, $\sigma_T^2$,
for the same values of $p$.}
    \label{fig:beta_vs_varT_multipanel}
\end{figure*}

To relate the transport response to the edge disorder, we consider the
mean and variance of the binary transmission distribution(see equation 25),

The mean transmission characterizes the average bond transmission, while
the variance measures the heterogeneity of the transmission values.
Figure~\ref{fig:beta_vs_varT_multipanel} compares the mean transport
exponent with the transmission variance as $f$ varies. The variance
vanishes at both $p=0$ and $p=1$, since the transmission is uniform across
the bonds in either limit. These limits nevertheless correspond to
different bond configurations: at $p=0$ all bonds are pristine, whereas
at $p=1$ all bonds are defective and transmit with probability $f$.
Thus, a vanishing transmission variance does not imply that the lattice
is free of defects.

\begin{figure*}[!t]
    \centering
    \includegraphics[width=0.94\textwidth,height=0.42\textheight,keepaspectratio]{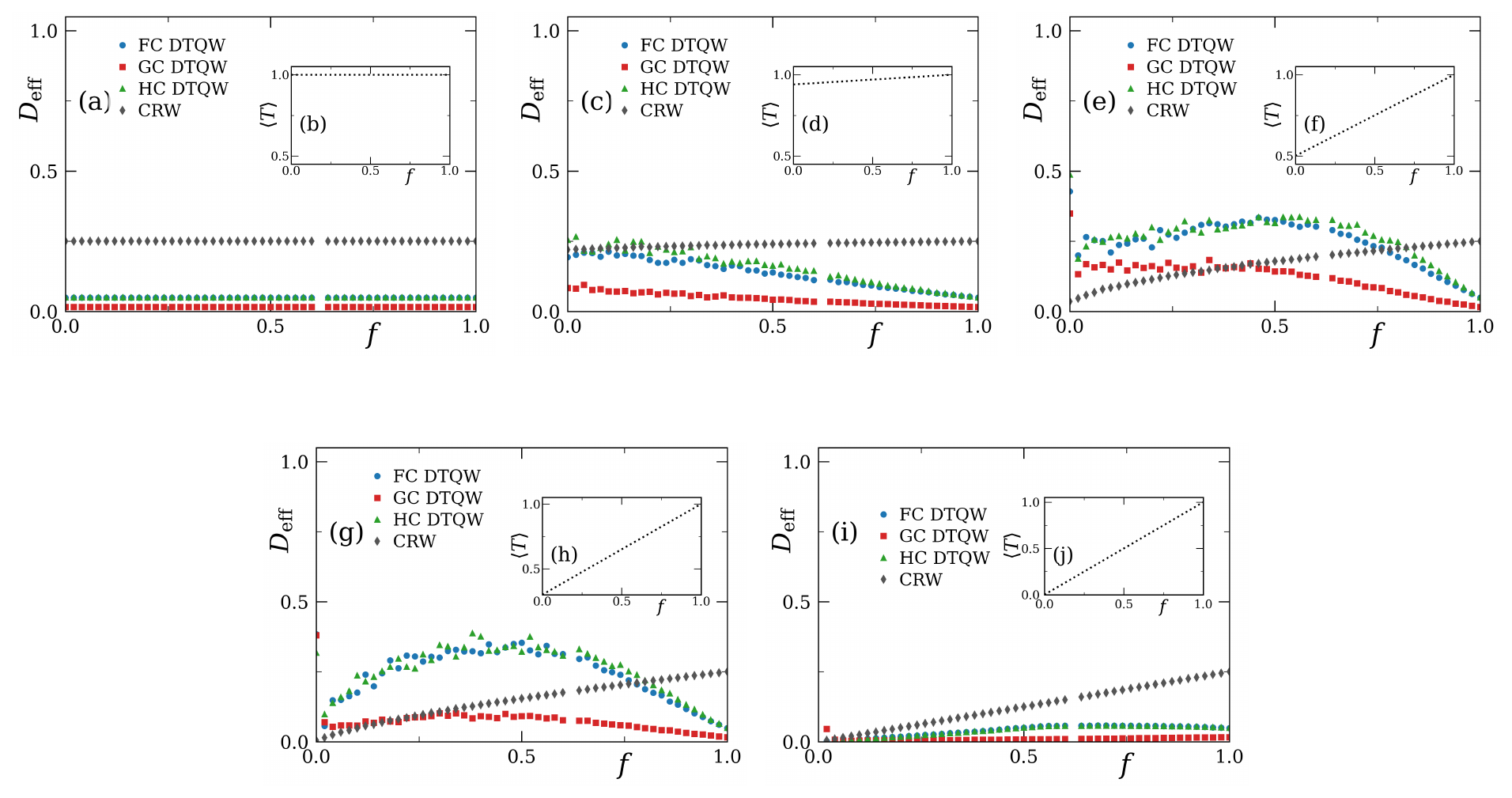}
    \caption{Effective diffusion coefficient $D_{\mathrm{eff}}$as a function
of the transmission probability $f$ for defect probabilities
$p=0$, $0.06$, $0.50$, $0.70$, and $1.00$ in panels (a), (c), (e),
(g), and (i), respectively. The corresponding panels (b), (d), (f),
(h), and (j) show the average of bond transmission, $\langle T\rangle$,
for the same values of $p$..}
    \label{fig:moments_and_deff}
\end{figure*}

Figure~\ref{fig:moments_and_deff} compares the effective diffusion
coefficient with the mean transmission. Together, these figures test
whether simple statistics of the bond transmissions track the measured
transport response. Differences between the classical and quantum walks,
as well as differences among the quantum coin operators, indicate that
the mean and variance alone do not provide a complete description of the
observed dynamics. The quantum response may depend on coherent
propagation in addition to the statistical properties of individual bonds.

\subsection{Time-Dependent Spreading and Spatial Concentration}
\label{subsec:ipr_and_beta_dynamics}

The transport maps characterize the measured spreading rates but do not
show how the spatial probability distribution evolves over time. We
therefore examine two complementary observables: the inverse participation
ratio (IPR), which measures spatial concentration, and the local transport
exponent, which tracks the time-dependent spreading rate.

\begin{figure*}[!t]
    \centering
    \includegraphics[width=1.0\textwidth,height=0.43\textheight,keepaspectratio]{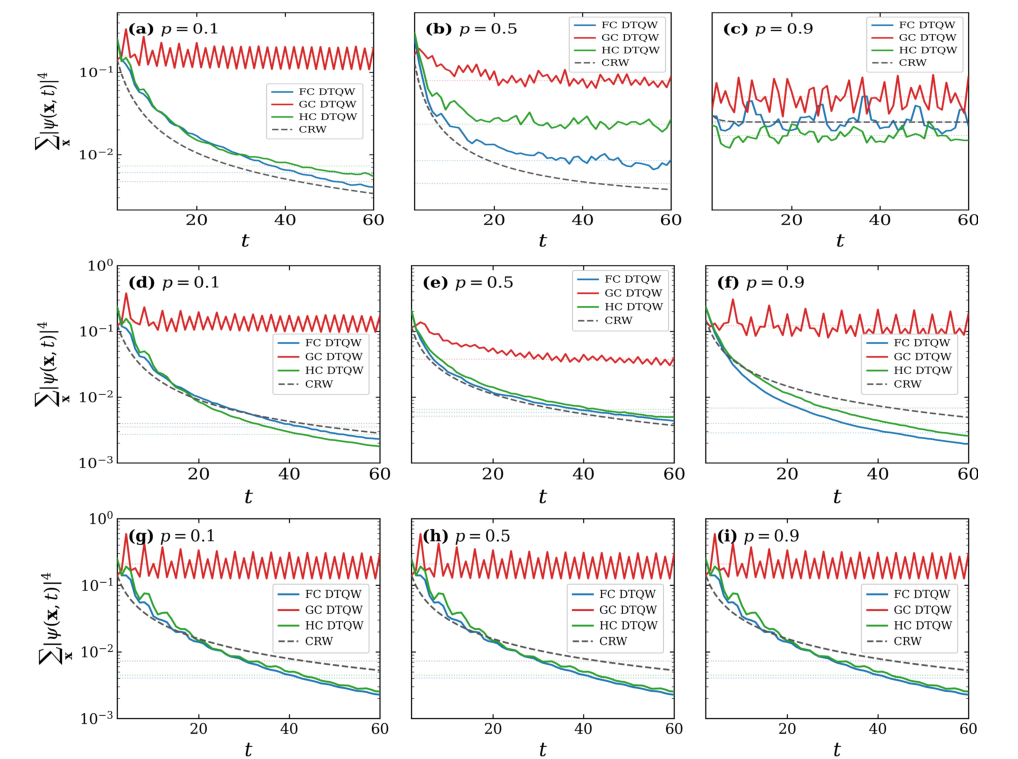}
    \caption{Time evolution of the spatial inverse participation ratio
    $\mathrm{IPR}(t)=\sum_{x,y}P(x,y,t)^2$.The three Rows correspond to
    $f=0.00,0.50,1.00$ from top to down and columns to $p=0.10,0.50,0.90$, from left to right. The panels
    compare IPR(t) for the classical and quantum walks over the simulated interval.}
    \label{fig:ipr_matrix_9panel}
\end{figure*}

Figure~\ref{fig:ipr_matrix_9panel} shows the evolution of the IPR for
representative values of $f$ and $p$. A plateau in the IPR indicates that
the spatial concentration changes little over the observed interval.
Figure~\ref{fig:beta_matrix_9panel} shows the corresponding local transport
exponents, allowing the time dependence of the spreading rate to be
examined alongside the IPR. If the local exponent decreases while the IPR
approaches a plateau, the two observables jointly support the
interpretation that spatial expansion is being suppressed over the
simulated interval. The extent of this suppression depends on the coin
operator and disorder parameters.

These finite-time signatures should be interpreted cautiously. Neither
an IPR plateau nor a decreasing local exponent alone establishes
asymptotic Anderson localization. Longer-time simulations and systematic
finite-size analysis would be needed to determine whether the observed
behavior persists asymptotically.

\subsection{Finite-Size and Fitting-Window Robustness}
\label{subsec:scaling_analysis}

\begin{figure*}[!t]
    \centering
    \includegraphics[width=0.90\textwidth,height=0.40\textheight,keepaspectratio]{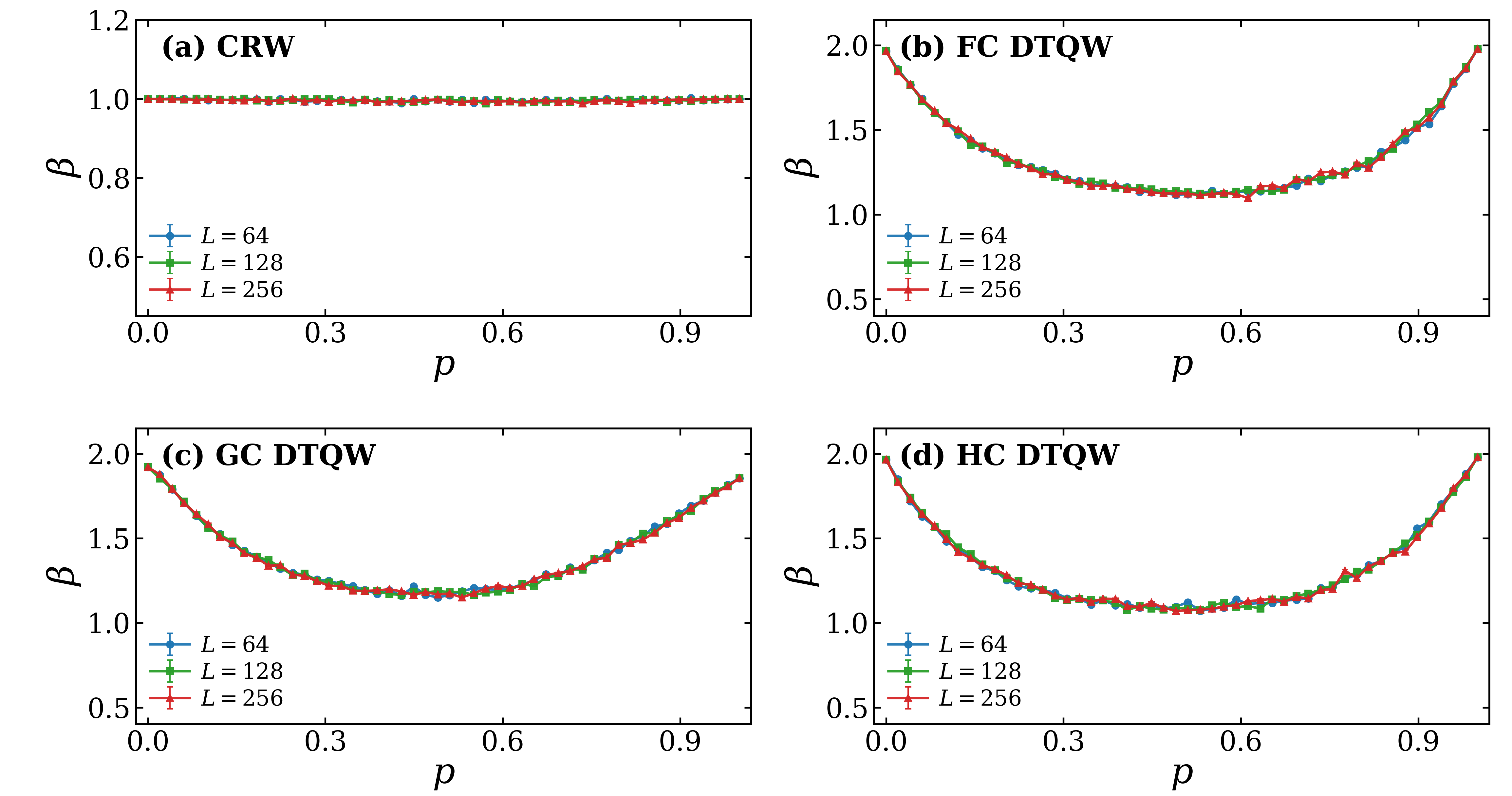}
    \caption{Finite-size comparison of the transport exponent for  classical walk (a) and the Fourier (b), Grover (c), and Hadamard (d) quantum walk, on disordered square lattices of size $L=64$, $128$, and $256$ with $f=0.50$, fitted for t=15 to t=30.}
    \label{fig:finite_size_scaling}
\end{figure*}

The transport exponents are estimated over a finite time window, so it
is important to test whether the results are sensitive to the lattice
size. Figure~\ref{fig:finite_size_scaling} compares the exponent for
L=64,128,256 at $f=0.50$, in the timing window $t=15$ to $t=30$.
The degree of agreement among the estimates indicates how strongly the
results depend on system size under these conditions. If the curves agree
closely, the estimates can be regarded as not strongly size-dependent
within the tested window.

This comparison is a finite-size consistency check, not proof of
thermodynamic-limit behavior or asymptotic localization. The conclusions
remain restricted to the system sizes, parameter values, and fitting
window examined here. Testing additional time windows and extending the
simulations would provide a stronger assessment of the robustness of
the measured exponents.

\FloatBarrier
\section{Conclusion and Outlook}
\label{sec:conclusion}

In this work, we presented a comprehensive numerical and comparative investigation of transport dynamics and localization phenomena in DTQWs and CRWs across disordered two-dimensional square lattices. By contrasting the Fourier, Grover, and Hadamard coin operators against classical diffusion, our show the fundamental differences between classical stochastic transport and coherent wave propagation in disordered low-dimensional media.

A central insight guiding our framework is the methodological motivation for transitioning from strong bond percolation to partial transmission edge disorder. Under standard binary bond percolation, where lattice edges are completely removed with probability $1-p$, structural link absences create severe topological ruptures that dominate the transport degradation. In this strong-disorder regime, both classical diffusion and quantum walks experience an essentially identical collapse in transport efficiency.So, this case completely erase the distinct signatures of coin operator symmetries. To circumvent this topological obliteration and investigate how intrinsic coin symmetries and coherent phase interference govern transport resilience, we implemented the partially transmitting edge lattice model. This model preserves underlying lattice connectivity while introducing controlled local phase shifts and scattering, thereby uncovering sharp, coin-dependent transport regimes that remain entirely obscured in conventional percolation models.

Our dynamical diagnostics provide direct numerical evidence for the absence of sustained quantum diffusion in two-dimensional non-interacting lattices subject to weak static disorder. By isolating early-time bulk propagation within the causal light-cone window prior to boundary reflections, we revealed contrasting dynamical scaling between classical and quantum walks. Whil CRWs asymptotically settle into a stabilized diffusion exponent ($\beta \approx 1$ or fixed subdiffusive values), quantum walks exhibit a persistent, monotonic decrease in $\beta(t)$ past the microscopic transient regime. This decaying scaling exponent, coupled with the asymptotic saturation of the inverse participation ratio ($\mathrm{IPR}$), demonstrates that coherent wave packets do not support an intermediate, stable quantum-diffusive phase; rather, cumulative destructive interference drives the system toward Anderson localization, aligning with the one-parameter scaling theory in two dimensions.

Furthermore, mapping the non-equilibrium phase diagrams across the $(f, p)$ parameter space established the critical dynamical interface at $\beta = 1$, which separates domains of transient super-diffusive mobility from destructive sub-diffusive arrest. We demonstrated that the position and peak of this interface depend strongly on the spatial heterogeneity of the disorder and the specific coin symmetry. The Grover coin operator, while sustaining higher early-time effective diffusion coefficients $D_{\mathrm{eff}}$ due to preferential diagonal transmission, ultimately localizes into a more spatially confined volume than the Hadamard or Fourier coins. Statistical analysis against network transmission moments highlighted that while classical transport correlates monotonically with transmission variance $\sigma_T^2$, quantum propagation decouples from simple disorder moments, governed instead by multi-path interference pathways.

These results hold direct experimental relevance for state-of-the-art quantum simulation platforms. Two-dimensional photonic waveguide arrays, coupled optical resonator networks, and multi-core optical fiber setups can implement both phase-tuned beam-splitter defects and variable coin operations with high fidelity. Our finite-size scaling analysis establishes that the onset of $\beta(t)$ decay and IPR saturation occurs within moderate temporal and spatial scales well before boundary reflection limits, rendering these dynamical transitions directly observable in existing laboratory-scale photonic architectures.

Some possible avenues for future exploration arising from this work are:

{\tiny $\bullet$}  { Synthetic gauge fields and spin-orbit couplings:} Introducing synthetic magnetic fields or artificial spin-orbit coupling into the coin operations can break time-reversal invariance or alter the underlying symmetry class (e.g., transition from orthogonal to symplectic ensembles), potentially stabilizing critical states or inducing delocalization transitions in two dimensions.

{\tiny $\bullet$}   {Interacting multi-particle walks:} Extending this framework to two-particle quantum walks with on-site bosonic or fermionic interactions would provide valuable insight into how spatial entanglement and exchange statistics compete against disorder-induced Anderson localization in planar topologies.

{\tiny $\bullet$}   {Extentsion to more complex topologies:} Generalizing the beam-splitter scattering formalism to scale-free, small-world, and other complex  networks will help elucidate how interference-driven localization impacts structural robustness and coherent excitation transfer across non-Euclidean geometries.

\bibliography{references}

\end{document}